# Blockchain 3.0 Smart Contracts in E-Government 3.0 Applications

**Sofia Terzi** [1]
CERTH/ITI
Aristotle University of Thessaloniki
sterzi@iti.gr
sofia.terzi@csd.auth.gr

**Konstantinos Votis** [2]
CERTH/ITI
kvotis@iti.gr

**Dimitrios Tzovaras** [3]
CERTH/ITI
dimitrios.tzovaras@iti.gr

**Ioannis Stamelos** [4]
Aristotle University of Thessaloniki
stamelos@csd.auth.gr

**Kelly Cooper** [5]
University of Maryland Global Campus
kelly.cooper@faculty.umuc.edu

*Abstract*—The adoption of Information Communication Technologies (ICT) and Web 3.0 contributes to the e-government sector by transforming how public administrations provide advanced and innovative services to interact with citizens. Blockchain (BC) and Artificial Intelligence (AI) disruptive technologies will reshape how we live, work, and interact with government sectors and industries. This paper presents how Blockchain 3.0 and Artificial Intelligence enhance robust, secure, scalable, and authenticity provenance solutions. Two validation scenarios are analyzed to present how blockchain smart contracts and AI agents support energy and health-oriented e-government services.

## Introduction

Blockchain (BC) technology is recognized as a critical, disruptive technology for many industries and applications. Starting with Bitcoin [7], a finance-oriented extremely ingenious distributed shared ledger and peer-to-peer value transfer technology, BC established trust between unknown stakeholders and automated payments. Bitcoin reformed the finance and supply chain industry by shortening the time needed to complete time-consuming processes and removing nearly all intermediaries.

Blockchain technology for financial payments automation without intermediaries is known as Blockchain 1.0. The technology acknowledged as Blockchain 2.0 followed with the Ethereum project [8], which differed from BC 1.0 with its support for smart contracts (SC). Other BC 2.0 technology projects include Hyperledger's HL Fabric, Sawtooth, Iroha [9], and R3's Corda [10]. Smart contracts (SC) are computer programs written to run on a blockchain and provide security and automation systems, making it possible for participating parties to agree upon certain conditions and actions to be performed when the conditions are met. These features of SCs reshape supply chain processes by enabling additional on-chain actions such as assets tracking and, in parallel, equip BCs with necessary characteristics for business cases outside of the supply chain. Blockchain is now used in industries such as healthcare [1][2], education [3], government [4], charities [5], real estate [6], insurance [16], and banking [15]. This expanded field of applications supported by BC is called Blockchain 3.0 because solutions are not restricted to finance actions and assets transfer [18] [19]. With the rise of Blockchain 3.0 technology, based on Directed Acyclic Graph (DAG) data structures [39], BC systems are more efficient, scalable, highly interoperable, and offer a better user experience. Among these sectors, government use cases are of special interest due to the implications they introduce when adopting a BC infrastructure. These implications may include internal issues related to a government such as politicians' inaction and opposition, or external issues such as digital transformation laws and sensitive citizens' and civil servants' personal data [17]. The BC's characteristics of decentralization provide zero down-time, ensure tamper-proof data and non-repudiation with immutability, implement security with cryptography to establish trust between participating parties, and utilize consensus algorithms for data integrity, verification, and scalability on private and permissioned blockchains [20].

Blockchain 3.0 technology supports the evolution for EG to become Web 3.0 oriented by providing the infrastructure, services, and processes needed alongside Information and Communication Technologies [21] such as Artificial Intelligence (AI) agents to secure and enhance communication between governments, businesses, and citizens [22]. EG 3.0 is totally dependent on Information Communication Technologies (ICT) to evolve along with Web 3.0 technologies, such as blockchain, artificial intelligence, semantic web and text analytics, machine learning, internet of things, and big data analytics [23].

This paper examines BC 3.0 and SC characteristics and features expected to contribute to EG 3.0 applications and offers selected best practices for how to incorporate BC 3.0 and SC into the design and implementation of ICT Web 3.0 e-government solutions.

## Blockchain

The two major forms of blockchain implementations are public permissionless and private permissioned. The following sections present their most important characteristics regarding EG 3.0.

## Permissionless Blockchains

Permissionless BCs were the first generation of Distributed Ledger Technology (DLT) to provide decentralized ledgers as opposed to centralized databases. Bitcoin and Ethereum are the most known representatives of permissionless BCs. Their premise is that all transactions are transparent to every participant and are written on the ledger only after a consensus of the majority of peers is achieved. Each participant shares an identical copy of this data, called state, that is formed of blocks connected to each other through cryptographic hashes. This architecture makes it almost impossible to change or trick others about the data state or take advantage of assets exchanged or discarded without notice by other peers. A disadvantage of permissionless blockchains is they do not support any control over who enters or leaves the network. This lack of control can be detrimental for security and may lead to energy-draining and time-consuming block generation techniques [11] to enforce security. The potential side effects of block generation include system scalability and speed.

Permissionless BCs can be ideal for EG 3.0 applications when data must be public and transparent. Such use cases may include the education sector verified and shared certificates, degrees, and diplomas issued by governmental organizations and academic institutions [40][41]. Other use cases include publishing voting results and disseminating publicly available documents and copyrights.

## Permissioned Blockchains

Due to BC's unique characteristics of immutability and decentralization, blockchain technology evolved beyond BC 1.0 to business priorities such as asset tracking and logging, consent and agreement enforcement and monitoring, and identity authentication and authorization. Permissionless blockchains achieve a great deal of decentralization; however, they can not guarantee the privacy and safety needed for sensitive citizen and government data. The lack of control over permissionless BCs and the exit and entry of network participants make documents, records, historical data, and transactions containing citizens' data visible.

Permissioned blockchains, such as Hyperledger (HL) Fabric, answer the need for private, decentralized, secure, and verifiable transactions among governments, citizens, and businesses. Although all transactions are written through smart contracts to the ledger, as they are in BC 1.0, permission must be given to access any data. On permissioned BCs, participants are strictly controlled by a central authority. In an EG use case, this may be a ministry or an independent authority. Blockchain policies exist on the network to grant permission to stakeholders to perform specific actions. For example, a citizen must be informed that a public administration organization requests specified data and the citizen must consent for access to be granted. These requests and consent actions are written on the blockchain to provide transparency for participants. Permissioned BCs address the need for privacy, scalability, security, and speed, although compromises are made in decentralization. When a central authority is introduced to authorize the private network's participants, decentralization is hindered and a BC controlling authority accesses the network [12].

Permissioned BCs are ideal for governmental applications that require a level of security such as an internal exchange of documents among public organizations for inventories, registries, or other private records.

## Smart Contracts

Smart Contracts (SC) [13] are computer programs immutably written on the blockchain and called by BC participants. SCs provide automation and control flow logic to any system BCs support. Smart contracts must be treated as software functions in every aspect and smart contract BC engines must be deterministic. The determinism of SCs is the characteristic that maintains the ledger at a stable, consistent state, enforces transaction finality, and avoids soft and hard forks [14]. The determinism of SC's actions is usually left to the developer. Thus, she must ensure automated actions are executed as planned and the results of these actions leave the data in a consistent state, regardless of the node(s) they are executed on. SC's actions must achieve the same result each time the SC is executed. In the writers' opinions, derived from empiricism, smart contracts can be categorized into three major categories:

- Static
- Dynamic
- Oracle driven

Depending on the specific use case to be implemented, the developer designs either dynamic, static, or oracle driven smart contracts. A definition of each, below, explores their characteristics to assist researchers, architects, and developers as they determine which is appropriate per use case.

### Static *standard output*

Static SCs do not call other smart contracts, do not reside on human interaction, take place in one-step, and never change their predefined number of actions. Static SCs perform primitive math operations such as addition, subtraction, multiplication, and division. Other SCs can call, retrieve, and consume the results of their operation. All SCs receive parameters to perform actions and are somehow dynamic. However, there are no additional conditions embedded in static SCs to change their path of action. Math operations consistently reach the same result and operators follow the same precedence rules every time. SCs can return a "yes/no" response to a specific question or return a standard image when an action is triggered. An EG 3.0 application example is a function that accepts a verification request for an academic diploma, looks to the ledger for the diploma holder, issues the institution name and date of issuance, and returns the result to the requester.

### Dynamic non-standard output

Dynamic SCs embed various rules that allow them to perform different actions. Examples of dynamic SCs include functions that monitor certain conditions and trigger intended actions. For example, when a dynamic SC monitors electricity consumption and temperatures logged on the BC of an energy-smart building. The dynamic SC includes thresholds for heating and consumption measurements to adjust temperatures in an eco-friendly way designed to avoid excessive electricity consumption and cost. The following pseudocode offers the logic behind monitoring and execution:

```
if room_temperature < 18 Celcius {
  if electricity_consumption < 25 Watt
          then turn_on air_conditioner
  else send_message: "The daily
          electricity       consumption
          threshold has been reached.
          Would you like to turn on the
          A/C?"
    if user_answer == 'YES' then
            turn_on air_conditioner
    else do_nothing }
if room_temperature > 25 Celcius {
  if electricity_consumption < 25 Watt
        then turn_on heating_unit
  else send_message: "The daily electricity
      consumption threshold has been
      reached. Would you like to turn on
      the heating_unit?"
    if user_answer == 'YES' then
          turn_on heating_unit
    else do_nothing }
```

This dynamic SC, although deterministic, follows a non-static conditioned flow that shows how a dynamic SC might be formed and how it can act. The code is simplistic and computer functions can be long and complex. Additionally, the example involves human interaction which, on occasion, may hinder or cancel the dynamic action feature of the SC. Human input is considered dynamic in terms of a non-standard, condition-driven final action. The dynamic nature of SCs may be controlled with machine-to-machine (M2M) actions. Unpredictable outcomes may occur if a developer's design and implementation of the SC are erroneous, incomplete, or non-deterministic.

Another approach to dynamic SC EG 3.0 applications is to interconnect public administrations that request to exchange citizen data. For example, if a tax service requests access to citizen land titles held by a land registry service. A dynamic SC supplied with a tax service VAT number may access land titles tied to that VAT number, if appropriate citizen permissions are in place. If a universal BC ledger contains land titles for all citizens, a dynamic SC may help to confront fraud and tax evasion and mediate the secure exchange of data between nations.

**Oracle driven**

Both static and dynamic SCs handle data that resides on the BC. Oracle, the third major category of SC, is designed to work with data from sources external to the BC. Oracle SCs are dynamic and include information brought in by the so-called AI oracles, which are also smart contracts. OracleSCs act as AI agents with the ability to request information from the real world and write it on the blockchain for other smart contracts to consume [24]. What is special about the oracle SC category is that SCs are generally not allowed to incorporate data external to the BC due to the determinism of BC functions. Determinism states the same result must be returned each time an SC function is called and external resources are often subject to change. Determinism is typically enforced by utilizing data that exists as the ledger's state. An exception is made through oracles to write data on the BC that represents the ledger state at the exact time the data is written on the ledger.

AI oracle SCs apply EG 3.0 to law applications. For example, laws for inheritance can change and notaries or other public servants in an oversight role must be formally informed regarding issues such as legacy transfer. An AI oracle accesses information from a government repository and writes to the BC when a specific law changes. After this, a notification is sent through a BC 3.0 application to prove date and time sent, to inform interested parties, and to request and record confirmation of receipt on the BC.

## E-Government 3.0

EG, by [25] definition, is the use of ICT to provide a means for governments, citizens, and businesses to interact, communicate, share information, and deliver services to various stakeholders. EG 1.0 utilized the World Wide Web and available ICTs to strive toward efficiency [26]. EG 2.0, through portal services supported by Web 2.0 technologies, became more citizen-centric, promoting citizen participation and enhancing e-democracy [27]. The technological evolution shaping EG infers EG 3.0 will use Web 3.0 ICTs such as distributed ledger technology (DLT), AI, Semantic Web, and the World Wide Virtual Web [20][28].

Artificial Intelligence is a promising and disruptive technology. AI's technological ability to equip machines with cognitive capabilities that learn, infer, and adapt per consumed data is reinforced by the amount of information produced by smart devices, social media, and web applications [29]. One problem governments, organizations, and companies face in leveraging this amount of information is centralization and provenance, the latter related to information source legitimacy and authenticity. Data in AI projects are centrally controlled and can be tampered with. For example, Microsoft's AI Twitter-based bot project was overwhelmed with racist remarks which, unfortunately, bots repeated to users [30].

One argument under consideration [22] offers AI as the solution to major governmental obstacles, particularly related to issues such as resource allocation, large datasets, experts shortage, predictable scenarios, procedural and repetitive tasks, and diverse data aggregation and summarization. Crucial to research is an analysis of how to overcome centralization, provenance, and authenticity problems. The combination of BC and AI technologies address current centralization problems and, in parallel, provide solutions for resource optimization and return private, personal data control back to their respective owners in a distributed, decentralized, and democratized manner [31].

The remainder of this paper examines two EG 3.0 scenarios supported by BC 3.0 and AI technology, the purpose of which is to provide EG stakeholders and policymakers avenues to exploit current industry BC and AI applications for governmental, public, and social good.

**Energy data – Scenario 1**

In recent years, digital smart city governance with ICT expanded and regional research addressed the increased energy demand that emanated from the multiplication and complexity of Internet of Things (IoT) devices. It became crucial for local governments to practice energy management strategies and use available energy efficiently [32]. A modern smart city applies smart technologies to its infrastructures and to citizen residences. The EG 3.0 scenario includes IoT devices, installed at citizen residences, that produce energy; these citizens are referred to as prosumers. This ability of energy consumers to produce energy from renewable sources and distribute that energy, through smart grids as prosumers, increases the difficulty of national energy management. However, prosumers also create the opportunity for smart city energy sustainability and efficiency when citizen produced energy is successfully modeled and incorporated into city energy systems along with energy related to transportation and facilities [33]. Energy management is critical; the European Commission, in the last two years, published two directives for energy efficiency goals with a 20% energy savings target by 2020 and a 30% energy efficiency target for 2030. Additional, specific national targets include lowering energy bills, reducing nation' reliance on external suppliers, and eco-friendly protecting the environment [42][43]. EG 3.0 supports citizen-sourcing, increases efficiency in all phases of the energy supply, and leads energy sector management. BC 3.0 technology, in conjunction with AI, provides authentication, decentralized intelligence, security, and collective decision making.

In EG 3.0, IoT devices produce energy data that is stored on a private permissioned blockchain. Data stored on a BC is tamper-proof; it is cryptographically immutable and authenticated because each transaction is digitally signed. Energy data is considered confidential, security concerns must be mitigated with a private permissioned BC. Know Your Customer (KYC) compliance is enforced through permission policies on the BC network; each citizen determines what personal information or energy production data is shared. Additional security is realized when prosumer registration applicants follow a strict protocol and participate in local energy networks logged on the BC. This prosumer energy approach is automated with Dynamic SCs controlling the processes of IoT data logging, registration and approval logging, and available energy dispatching and monitoring.

A SC collects energy consumption and production measurements from prosumer IoT devices and logs them on the blockchain. The prosumer provides the BC login identity issued to her. This self-sovereign identity (SSI) ensures secure entry and prosumer user control [44]. A SC dispatches surplus energy from a prosumer residence to the main energy system or to a citizen-sourced smart grid. If, for example, daily consumption need is 14kWh and the SC detects the power produced from renewable sources exceeds 15kWh, an action automatically triggers and dispatches this surplus power to a pre-identified local energy system. A dynamic smart contract deposits the required, predefined payment for the energy dispatched to the prosumer's account. Oracle-based SCs inform citizens how their energy dispatching can be more profitable and provide incentives for participants on the energy network. Smart grids, informed by local policies, consider geographic factors, energy needs, and building production capabilities. AI agents operate at citizen residences as collective decision-making mechanisms that apply Swarm Intelligence (SI) and achieve swarm goals [34]. SI calculates how much energy can be dispatched to a city's central energy system and how much energy is available to be traded among smart grid participants. AI EG applications read data written on the BC and forecast city energy needs for hours, days, or months. AI analyzes data for trends or peak hours. The results and metadata from AI analyses are grouped per district to help governments and policymakers create more efficient energy management strategies as they achieve local and national goals.

**Health data – Scenario 2**

National Healthcare systems are a sector where e-health strategies must be adopted for governments to control excessive healthcare costs [35]. Healthcare systems hold massive amounts of confidential data; problems arising in processing and analyzing this big data are solvable with AI. Research shows older adults struggle to use e-health systems [36] [37]. With AI chatbots speech recognition support older adult questions and inquiries; chatbots can provide responses and guidance. AI agents also support patient forms completion and submittal to appropriate government departments [38]. A permissioned BC 3.0 secures the confidentiality and authenticity of private e-health data. EG 3.0 provides solutions to e-health priorities by utilizing ICT and Web 3.0 to transform legacy systems, increase their efficiency and effectiveness, decrease costs, and provide citizen-centric health care services [36].

The EG ecosystem includes IoT wearables that send patient data, such as heart rate or blood pressure, to a private, permissioned BC that ensures data security, authenticity, and confidentiality. When a doctor requests access to patient records and data, the BC triggers an SC and the action logs on the BC. The SC then forwards the doctor's request message to the patient and the patient approves data access for the doctor. The SC writes patient approval or denial on the BC. This way a complete tracking system for requests and consent responses is formed. This secure process applies to e-health records exchanged at national or international levels with intact end-to-end security.

## Further Research

We acknowledge restrictions apply in our research, mainly due to the different energy and e-health implementations among countries in Europe. Our research focuses on governments and citizens, and further research will include applications and results with public administrations and civil servants. The scenarios demonstrated focus on BC 3.0 support. Thus, EG scenarios that include additional Web 3.0 technologies must be designed, developed, and tested. We hope to contribute more on these subjects as our research projects progress.